\documentclass{article}
\usepackage{amsfonts,cite}

\newcommand{\be}{\begin{equation}}
\newcommand{\ee}{\end{equation}}
\newcommand{\bea}{\begin{eqnarray}}
\newcommand{\eea}{\end{eqnarray}}
\newcommand{\ba}{\begin{array}}
\newcommand{\ea}{\end{array}}

\newcommand{\al}{\alpha}
\newcommand{\ga}{\gamma}

\newcommand{\bet}{\beta}
\newcommand{\ka}{\kappa}
\newcommand{\de}{\delta}
\newcommand{\ep}{\epsilon}

\newcommand{\la}{\lambda}

\newcommand{\ze}{\zeta}

\newcommand{\Ups}{\Upsilon}

\newcommand{\Z}{\mathbb{Z}}
\newcommand{\R}{\mathbb{R}}

\newcommand{\im}{{\rm Im}}
\newcommand{\re}{{\rm Re}}
\newcommand{\D}{{\rm d}}
\newcommand{\DD}{{\rm D}}

\newcommand{\id}{\hbox{1\kern-.27em l}}
\newcommand{\sid}{\hbox{\scriptsize1\kern-.27em l}}

\newcommand{\pa}{\partial}
\newcommand{\rar}{\rightarrow}
\newcommand{\non}{\nonumber}
\newcommand{\we}{\kern-.1em\wedge\kern-.1em}
\newcommand{\scal}{\kern-.13em\cdot\kern-.13em}
\newcommand{\tF}{\tilde{F}}
\newcommand{\tA}{\tilde{A}}

\newcommand{\half}{\mbox{$\frac{1}{2}$}}
\newcommand{\third}{\mbox{$\frac{1}{3}$}}

\newcommand{\sixth}{\mbox{$\frac{1}{6}$}}
\newcommand{\II}{I\kern-.09em I}
\newcommand{\cC}{{\mathcal C}}
\newcommand{\cH}{{\mathcal H}}
\newcommand{\cF}{{\mathcal F}}
\newcommand{\cO}{{\mathcal O}}

\newcommand{\cU}{{\mathcal U}}
\newcommand{\cK}{{\mathcal K}}

\newcommand{\bcH}{\bar{\mathcal H}}
\newcommand{\bcF}{\bar{\mathcal F}}

\newcommand{\bcU}{\bar{\mathcal U}}

\renewcommand{\thefootnote}{\fnsymbol{footnote}}

\begin{document}

\vspace*{-12mm}
\rightline{\vbox{\small
\hbox{DAMTP-1999-156}
\hbox{NORDITA-1999/68 HE}
\hbox{\tt hep-th/9911052}
}}
\bigskip


\begin{center}

{\Large \bf Super-$p$-brane actions from interpolating 
dualisations\footnote{Talk presented by A.W. at QFTHEP'99, Moscow, 
27 May--2 June 1999, and at the Third annual meeting of the European TMR
network {\em Quantum aspects of gauge theories, supersymmetry and 
unification}, Paris, 1--7 September 1999.}}\\ 

\vspace{4mm}


{\large Anders Westerberg$^{a,b}$ and Niclas Wyllard$^{a}$}\\[6pt]
$^a$DAMTP, University of Cambridge, Silver Street, Cambridge CB3 9EW,
United Kingdom\\[3pt]
$^b$NORDITA, Blegdamsvej 17, DK-2100 Copenhagen, Denmark\footnote{Present
address.}

\end{center}


\begin{abstract}
\noindent
We review a recently proposed method for constructing super-$p$-brane 
world-volume actions. In this approach, starting from a democratic choice of 
world-volume gauge-fields guided by $p$-brane intersection rules, 
the requirements of $\kappa$-symmetry and gauge invariance can be used to 
determine the corresponding action. 
We discuss the application of the method to some cases of interest, 
notably the $(p,q)$-5-branes of type-{\II}B string theory in a
manifestly S-duality covariant formulation.
\end{abstract}

\setcounter{footnote}{0}\renewcommand{\thefootnote}{\arabic{footnote}}


\subsection*{Introduction}

With the derivation of the $\kappa$-symmetric actions for the 
D$p$-branes~\cite{Cederwall:1996b,Cederwall:1996c,Aganagic:1996a,
Aganagic:1996b,Bergshoeff:1996c} and the 
M5-brane~\cite{Bandos:1997,Aganagic:1997b}, 
the programme of constructing world-volume actions for all the points on the
`revisited brane-scan' of ref.~\cite{Duff:1993} was essentially completed by
early 1997\footnote{For the M5-brane case, note, however, the caveat
discussed in ref.~\cite{Witten:1996a}.}. 
However, the case of the NS5-brane of the type-{\II}B theory was not 
satisfactorily resolved by these efforts, and, moreover, since then a 
number of more or less `exotic' extended objects 
have been predicted from a detailed analysis of the 
relevant supertranslation algebras~\cite{Hull:1997} and/or U-duality 
considerations (see ref.~\cite{Obers:1998} and references therein). 
These facts have provided motivation for subsequent work on completing and 
refining the formulations of the world-volume dynamics for extended objects 
in string and M~theory. 

One line of further development has been the construction of manifestly 
S-duality covariant actions for the BPS multiplets of branes in the 
type-{\II}B theory, in particular for the dyonic 
$(p,q)$-strings~\cite{Cederwall:1997} and for the SL(2)-singlet 
D3-brane~\cite{Cederwall:1998a}. This programme, whose extension to the 
dyonic $(p,q)$-5-branes was given in ref.~\cite{Westerberg:1999b} and will 
be reviewed below, is based on a particular type of action in which 
the brane tension is generated by an auxiliary world-volume gauge field. More 
specifically (leaving the issue of SL(2) covariance aside until later), 
the actions we will consider are of the generic form
\begin{equation}
\label{action}
S = \int\D^{p+1}\xi\sqrt{-g}\,\la\,[1 + \Phi(\{F_{i}\}) - ({*}F_{p+1})^2]\,,
\end{equation}
where $\{F_{i}\}$ and $F_{p+1}$ denote a set of gauge-invariant abelian 
world-volume field strengths, each of the form 
$F= \D A - C - \{\mbox{composite terms}\}$; we generically use $C$ to denote 
(the superfield extension of) a supergravity gauge potential (here pulled back
to the world-volume). Moreover, the composite terms are wedge products of 
lower-rank $F$'s and $C$'s required for gauge invariance.
As indicated, the $(p{+}1)$-form field strength $F_{p+1}$ plays a special 
role: the equation of motion for its non-propagating gauge potential $A_p$ 
has the solution $\la\,{*}F_{p+1} = \mbox{constant}$, a constant which
can be identified with the tension. In addition to $A_p$, the action
contains a second auxiliary field, $\la$, which serves as a Lagrange 
multiplier for the constraint 
$\Ups=1 + \Phi({\{F_i\}}) - ({*}F_{p+1})^2 \approx 0$.\footnote{As
a special case, restricting the world-volume gauge-field content to the
Born--Infeld field strength $F_2=\D A_1-B_2$ (in addition to the tension 
field-strength) leads to an alternative formulation of the 
D-branes~\cite{Bergshoeff:1998c}. The standard formulation,
in which the action consists of the sum of a DBI and a WZ term, 
can be regained by eliminating $A_p$ and $\la$.}

In general, the choice of brane is determined by the leading background
gauge potential $C_{p+1}$ appearing in the tension field strength,
while the remaining world-volume gauge fields correspond to other branes
that can end on or intersect the world-volume. 
As we shall see, it is possible to construct covariant actions of the 
type~(\ref{action}) in which world-volume potentials appear together with 
their Poincar\'e duals; the problem of preserving the correct 
degree-of-freedom counting can be resolved in a rather elegant and natural way.
An original motivation for considering such formulations---beside their 
potential usefulness in studies of target-space brane configurations from a 
world-volume perspective---stems from the observation~\cite{Townsend:1996a} 
that relating the world-volume actions for branes which transform into each 
other under S-duality involves performing Poincar\'e-duality transformations 
on the world-volume gauge fields. As a recipe for constructing new brane 
actions 
from known ones, however, the latter approach has its limitations; the basic 
problem is that in order to express an action $S[F]$ in terms of the 
Poincar\'e-dual field strength $\tF\sim\frac{\de S[F]}{\de F}$, one has to 
solve an algebraic equation whose order increases with the world-volume 
dimension. In practice, this obstacle has, e.g., prevented the construction 
of a $\ka$-symmetric action for the type-{\II}B NS5-brane from the D5-brane 
action by means of a combined background ${\Z}_2$ S-duality transformation and
a world-volume Poincar\'e transformation of the Born--Infeld field strength 
$F_2$.\footnote{With motivation coming from T-duality considerations of
type-{\II}A KK monopoles, it has been suggested~\cite{Eyras:1998} that such a 
Poincar\'e-duality transformation is not needed in this particular case.}
By employing a formulation where both field strengths appear simultaneously
we are able to circumvent these problems, obtaining in the process a 
continuous family of actions interpolating between Poincar\'e-dual limits.

As it turns out, this method can be systematised to the extent that it becomes
an essentially algorithmic procedure for constructing $\ka$-symmetric 
world-volume actions. Led by the available background gauge potentials,
one constructs a set of gauge-invariant world-volume field strengths.
Except for the tension field strength, they all come in world-volume dual
pairs and each such pair has an associated interpolation parameter.
Inserting these field strengths in an Ansatz of the form~(\ref{action}), the 
requirement of $\ka$-symmetry turns out to be sufficient to determine 
the action. 
Before we go on to discuss more complicated applications, 
we shall first illustrate the method using a simple and transparent example;  
relevant references are the papers~\cite{Cederwall:1998a,Cederwall:1998b,
Westerberg:1999a,Westerberg:1999b} where more detailed accounts can be
found.

\subsection*{A simple example: the M2/D2-brane}

We consider the well-known case of the S-dual pair of the M2- and type-{\II}A 
D2-branes,\footnote{Actually, we will be working exclusively in $D=10$, so by 
the M2-brane we here mean the brane obtained by direct dimensional reduction 
on $S^1$ of the $D=11$ supermembrane. In other words, we consider the 
situation where the S-duality transformation from the eleven-dimensional
background theory down to ten dimensions has already been performed.} 
omitting some more technical points; for more details see 
ref.~\cite{Westerberg:1999b}. 
Let us begin by recalling some relevant facts about type-{\II}A supergravity. 
Among other fields, the bosonic sector of this theory contains the 
gauge-invariant field strengths $R_2$, $H_3$ and $R_4$ (the subscript 
indicating the form degree) which satisfy the Bianchi identities 
\be
\D R_2 = 0\,, \qquad \D H_3 = 0\,, \qquad \D R_4 = H_3\, R_2\,.
\ee
The first two relations imply that $R_2 = \D C_1$ and $H_3 = \D B_2$, 
whereas the third one allows for some freedom in defining $R_4$; 
more precisely, 
$R_4 = \D C_3 + x\,B_2 \, R_2 - (1{-}x)\,C_1\, H_3 = \D (C_3 + x\,C_1\, B_2) 
- C_1\, H_3$, 
the parameter $x$ thus arising from an ambiguity in the definition of $C_3$ 
corresponding to the field redefinition $C_3 \rar C_3 + x\, C_1 \, B_2$. 
Although changing the value of $x$ does not change the physics, the 
effect on the form of the world-volume theory can be significant.

As mentioned above, gauge invariance is one of the guiding principles in the 
construction of the actions. At the supergravity level, the background field 
strengths are invariant under the gauge transformations 
\be
\label{backgaug}
\de C_1 = \D L_0\,, \qquad \de B_2 = \D L_1\,, \qquad
\de C_3 = \D L_2 - x\,R_2 \, L_1 - (1{-}x)\,H_3\, L_0\,.
\ee
Each of the background gauge potentials couple minimally to a world-volume 
gauge field, and the gauge transformations~(\ref{backgaug}) thus induce 
transformations on the world-volume. In order for the world-volume field 
strengths to be gauge invariant, the latter have to be cancelled by 
transformations of the world-volume gauge potentials, possibly after the 
inclusion of sub-leading composite terms. For the case under consideration, 
the outcome of this analysis is the following set of 
world-volume field strengths:
\be
F_1 = \D A_0 - C_1\,, \qquad F_2 = \D A_1 - B_2\,, \qquad
F_3 = \D A_2 - C_3 + x\, B_2\, F_1 - (1{-}x)\, C_1\, F_2\,.
\ee
These field strengths are gauge invariant and satisfy the Bianchi identities
\be
\label{BIF123}
\D F_1 = -R_2 \,, \qquad \D F_2 = -H_3\,, \qquad
\D F_3 = -R_4 + x\,F_1 \, H_3 - (1{-}x)\,F_2 \, R_2\,.
\ee
Thus, although essentially trivial at the supergravity level, the field 
redefinition parameterised by $x$ has a significant impact on the formulation 
of the world-volume theory, as can be seen by the $x$-dependent expression for
the tension-form field strength $F_3$. For $x=0$, $F_1$ decouples and the 
sub-leading terms in $F_3$ can be recognised as the negative of the 
WZ-form in the D2-brane action, while for $x=1$ the tension field strength is
the one appropriate for the description of the dimensionally reduced M2-brane.
For intermediate values, on the other hand, $F_1$ and $F_2$ are 
present simultaneously, and consequently there is one bosonic degree
of freedom too many. Proceeding next to construct an action for the general 
case, we shall find that the resolution of this problem is inherent in the 
formalism~\cite{Cederwall:1998a}. We use the Ansatz\footnote{In this note
we will suppress the dilaton dependence; this dependence, which can be 
determined from the dilaton-scaling of the supergravity constraints,
is of importance when determining the $\ka$-variation of the action.}
\be
\label{D2ansatz}
S = \int \D^{3}\xi\sqrt{-g}\,\la \left[1 +
\Phi(F_1,F_2) - ({*}F_3)^2\right]\,.
\ee
The equation of motion for $A_2$ is $\D [\la\,{*}F_{3}]=0$, whereas those
for the dynamical gauge potentials $A_0$ and $A_1$ read 
\be
\label{eom}
\D\big[\la\{{*}\frac{\de\Phi}{\de F_1} + 2 x\, B_2 \,{*}F_3\}\big] = 0\,,
\qquad
\D\big[\la\{{*}\frac{\de\Phi}{\de F_2} - 2 (1{-}x)\,C_1\,{*}F_3\}\big] = 0\,.
\ee
By using the Bianchi identities for $F_1$ and $F_2$ together with the
equation of motion $\D[\la\,{*}F_{3}]=0$, the explicit dependence on the
background fields can be eliminated from these equations with the result
\be
\D\big[\la\{{*}\frac{\de\Phi}{\de F_1} - 2x\,F_2\,{*}F_3\}\big] = 0\,,\qquad
\D\big[\la\{{*}\frac{\de\Phi}{\de F_2} + 2 (1{-}x)\, F_1\, {*}F_3\}\big] = 0\,.
\ee
It is thus consistent with the equations of motion and the Bianchi
identities to impose the duality relations
\be
\label{D2dualrel}
-2x\,{*}F_3\,{*}F_2 = K_1 := \frac{\de\Phi}{\de F_1}\,,\qquad
2(1{-}x)\,{*}F_3\,{*}F_1 = K_2 := \frac{\de\Phi}{\de F_2}\,,
\ee
where $\Phi$ is yet to be determined. The effect of these relations is to 
reduce the total number of degrees of freedom contained in $A_1$ and $A_2$ by 
half, thus compensating for the doubling of gauge fields in the action. Since 
for the limiting value $x=0$ ($x=1$), $F_1$ ($F_2$) decouples, the 
degree-of-freedom counting works out correctly.
It is important to note that the duality relations must not be substituted 
into the action; rather, they supplement the equations of motion derived
from the latter.

The duality relations~(\ref{D2dualrel}) also play a crucial role in 
implementing $\ka$-symmetry---the other main guiding principle of the 
procedure---because they allow us to determine the variation of the 
action~(\ref{D2ansatz}) (or, equivalently, that of the constraint 
$\Ups=1+\Phi-({*}F_3)^2\approx0$) under the $\ka$-transformations
\bea
\label{basicvars}
\de_\ka g_{ij} &=& 2\,{E_{(i}}^a\,{E_{j)}}^B\ka^\al\,T_{\al Ba}\,, 
\quad
\de_\ka\phi= \ka^\al\pa_\al\phi \,, \quad
\de_\ka F_1 = -i_{\ka}R_2\,, \non\\
\de_{\ka}F_2 &=& -i_\ka H_3\,, \quad
\de_\ka F_3 = -i_{\ka}R_4+x\,F_1\,i_{\ka}H_3-(1{-}x)\,F_2\,i_{\ka}R_2\,,
\eea
in spite of the fact that we do not know what $\Phi$ is from the outset
(the background fields are now the superfield extensions of their
bosonic counterparts). 
To see how this works~\cite{Cederwall:1998b,Cederwall:1998a}, note that the 
scalar function $\Phi$ is formed out of contractions between the 
world-volume field strengths and the metric only,
and so a simple scaling argument shows that the variation of $\Ups$ can be 
written\footnote{The part of this expression proportional to the dilaton 
variation was derived by temporarily instating the dilaton-dependence of the 
action.}
\bea
\de_{\ka} \Ups &=& K^{i}\,\de_{\ka}F_{i} +
\mbox{$\frac{1}{2!}$}K^{ij}\,\de_{\ka}F_{ij} +
\mbox{$\frac{2}{3!}$}F^{ijk}\,\de_{\ka}F_{ijk} - \big[\half K^{(i}F^{j)} +
\mbox{$\frac{1}{2!}$}{K^{(i}}_{l}F^{j)l} \non\\ && +
\mbox{$\frac{3}{2}$}\scal \mbox{$\frac{2}{3!}$}{F^{(i}}_{lm}F^{j)lm} 
\big]\,\de_{\ka}g_{ij} +
\big[ \mbox{$\frac{3}{4}$}K_1\scal F_{1} -\half K_2\scal F_{2} 
+ \mbox{$\frac{2}{4}$} F_3\scal F_3\big]\,\de_{\ka}\phi\,.
\eea
By imposing the duality relations~(\ref{D2dualrel}), $K_1$ and $K_2$ get
replaced by explicit expressions in the world-volume field strengths.
Inserting the transformations~(\ref{basicvars}) with the supergravity on-shell 
constraints imposed on the background superfields and demanding that the 
irreducible components of the resulting variation vanish, then allows us to 
deduce the duality relations and subsequently the action. 
The final result is
\be
S = \int \D^3\xi\,\sqrt{-g}\,\la\,\big[1 + x\,F_1\scal F_1 + 
(1{-}x)\,F_2\scal F_2 +
x\,(1{-}x)\,F_1\scal F_1\;F_2\scal F_2 - ({*}F_3)^2\big]\,.
\ee
{}From this action one derives both the equations of motion 
(cf.~(\ref{eom})) and the supplementary duality relations 
(cf.~(\ref{D2dualrel})).
Another important object determined in the process is the $\ka$-symmetry
projection operator 
\be
2\,{*}F_3\,P_\pm = {*}F_3\,\id 
\mp\big[{*}\ga_3 - x\,{*}F_1\scal \ga_2\ga_{11} +
(1{-}x)\,{*}F_2\scal\ga_1 \ga_{11} \big]\,.
\ee

Let us at this point summarise the generic features of the procedure
which carry over to the more complicated cases to be discussed next:
starting from a world-volume gauge-field content motivated by
target-space considerations we have, by employing the principles of gauge
invariance and supersymmetry (at this level manifested as $\ka$-symmetry), 
found a one-parameter family of actions for the D2-brane interpolating between
two limiting Poincar\'e-dual cases. For intermediate parameter values the 
number of gauge fields in the action are doubled, the associated doubling of 
degrees of freedom being resolved by supplementing the equations of motion 
with duality relations inherent in the formalism.
and the supplementary duality relations

\subsection*{The Type-{\II}B NS5-brane}

On shell, the type-{\II}B NS5-brane---like its S-dual partner, the 
D5-brane---has $(1,1)$ supersymmetry in $D=6$ and is described by a vector
multiplet~\cite{Callan:1991}. Whether the $\ka$-symmetric action for the 
NS5-brane should contain a one-form gauge-potential (like the one for the 
D5-brane) or should be formulated in terms of a dual three-form potential
(as experience from other S-dual brane pairs would suggest) does however
not follow from this fact. Our formulation allows us to by-pass this question 
by introducing both potentials together with an associated interpolation 
parameter. These one- and three-form potentials, $\tA_1$ and $A_3$, can also 
be motivated by the fact (see, e.g., ref.~\cite{Argurio:1997}) that D-strings 
and D3-branes can end on the NS5-brane world-volume. Since the same goes for 
D5-branes, we also introduce an extra five-form gauge potential $A_5$, in 
addition to the tension potential $\tA_5$.

Using the additional fact that fundamental strings cannot end on the 
NS5-brane, we require that the Born--Infeld field strength $F_2$ be absent 
from the 
world-volume action. This assumption fixes the parameterisation of the
background field strengths, and we are led to RR background and world-volume 
field strengths that can be succinctly summarised in the expressions 
$R = e^{B}\,\D C$ and $e^{-B}F = \D A - C$, respectively. The latter 
expression is iterative and we have defined $F = F_0 + \tF_2 + F_4 + F_6$, 
where $F_0=\D A_{-1} - C_0$ ($\D A_{-1}$ formally denoting a constant) is 
dual to $F_6$ and needed for gauge invariance (Peccei--Quinn symmetry). 
In the same compact notation the world-volume Bianchi identities read 
\be
\D F = - R + H_3\, F\,.
\label{NS5wvBIs}
\ee
Furthermore, the gauge-invariant field strength of the NS-NS six-form gauge 
potential is chosen as 
\be
H_7 = \D B_6 - (1{-}y)\,C_6\,\D C_0 + y\,C_0\,\D C_6 - (1{-}x)\,C_2\, \D C_4 
+ x\, C_4\,\D C_2 \,,
\ee
an expression which preserves the Bianchi identity 
$\D H_7 = R_7\,R_1 - R_5 \,R_3$ and leads to 
a lengthy expression for the tension field strength $\tF_6$ (given in
ref.~\cite{Westerberg:1999a}) as well as a more compact one for its
Bianchi identity:
\be
\D \tF_6 = -H_7 -(1{-}y)\,R_7\,F_0 + y\,R_1\,F_6 - (1{-}x)\,R_3\,F_4 
+ x\,R_5\,\tF_2 
+ (1{-}x{-}y)\,H_3 \,F_0\, F_4 + (\half{-}x)\,H_3\,\tF_2\,\tF_2\,.
\label{NS5tF6BI}
\ee
Given the above field content, the Ansatz for the action (written in 
differential-form notation) takes the form
\be 
S= \int \la\, \big[{*}1+{*}\Phi(F_0,\tF_2,F_4,F_6)+\tF_6\,{*}\tF_6\big]\,,
\label{NS5ansatz}
\ee 
from which we can derive the following expressions for the duality
relations compatible with the Bianchi identities~(\ref{NS5wvBIs}):
\bea
\label{NS5dualrels}
2y\,{*}\tF_6\,F_0 = {*}K_6 := \frac{\de{*}\Phi}{\de F_6}\,, &\phantom{mm}&
-2(1{-}y)\,{*}\tF_6\,F_6={*}K_0:=\frac{\de{*}\Phi}{\de F_0}\,, \non \\
 -2(1{-}x)\,{*}\tF_6\,F_2={*}K_4:=\frac{\de{*}\Phi}{\de F_4}\,, &&
\phantom{-}2x\,{*}\tF_6\,F_4={*}K_2:=\frac{\de{*}\Phi}{\de F_2}\,.
\eea
Using the known on-shell constraints for type-{\II}B supergravity, one
can then proceed to determine the action by imposing $\ka$-symmetry.
The calculations, as well as the general results, are rather lengthy and 
can be found in ref.~\cite{Westerberg:1999a}. For certain values of 
the interpolation parameters, however, the expressions simplify considerably. 
For instance, when $x=\mbox{$\frac{2}{3}$}$ and $y=0$, $F_6$ decouples
and we have
\bea
2\,{*}\tF_6\,P_\pm &=& {*}\tF_6\,\id \pm [{*}\ga_6\,K + F_0\,{*}\ga_6\,J + 
\mbox{$\frac{2}{3}$}\,{*}\tF_2\scal \ga_4\,I \non \\
&& -\,\third\,\{F_0\,{*}F_4 - \half\,{*}(\tF_2 \we \tF_2)\}\scal \ga_2\,K
- \third\,{*}F_4\scal \ga_2\,J]\,,\\ 
{*}\Phi &=& F_0\,{*}F_0 + \mbox{$\frac{2}{3}$}\,\tF_2\we{*}\tF_2 
+ \third\,F_4\we{*}F_4  -\,\third\,F_0\,{*}F_4\we\tF_2\we\tF_2 
+ \third\,(F_0)^2\,F_4\we{*}F_4 \non\\
&&+\, \mbox{$\frac{2}{9}$}\,(\tF_2\we F_4)\,{*}(\tF_2\we F_4) 
+\, \sixth\,{*}({*}F_4\we{*}F_4)\we\tF_2\we\tF_2 
+\, \mbox{$\frac{1}{12}$}\,{*}(\tF_2\we\tF_2)\we\tF_2\we\tF_2\,.
\eea
{}From the latter expression the equations of motion and the duality relations 
may readily be derived.
Since the action is $\ka$-symmetric and couples correctly to the NS-NS 
six-form potential we can conclude that it describes the type-{\II}B NS5-brane.

\subsection*{Manifestly SL(2)-covariant actions}

As mentioned in the introduction, world-volume actions of the kind discussed 
above that are manifestly covariant under the SL(2,$\Z$) S-duality group can 
be constructed for the type-{\II}B 
branes~\cite{Cederwall:1997,Cederwall:1998a,Westerberg:1999b}.  
These constructions rely on the superspace formulation of type-{\II}B 
supergravity~\cite{Howe:1984} where the U(1) R-symmetry has been gauged,
allowing the scalars of the theory to transform linearly under SL(2,$\R$). 
More specifically, the scalars form a $2{\times}2$ matrix 
\be
\label{scalmatr}
\left(\ba{cc} \cU^1 & \bcU^1 \\ \cU^2 & \bcU^2 \ea \right)
\ee 
on which SL(2,$\R$) acts from the left and U(1) locally from the right, 
both group actions preserving the constraint
$\frac{i}{2}\ep_{rs}\cU^r\bcU^s = 1$ (here $\ep_{12}=-1$). 
The two physical scalars of the theory---the dilaton $\phi$ and the
axion $C_0$---are encoded in~(\ref{scalmatr}) as 
$\cU^1(\cU^2)^{-1}=-C_0+i\,e^{-\phi}$.
The SL(2) doublet $\cU^r$ links fields transforming in the fundamental
representation of SL(2) and SL(2)-invariant fields charged under the gauged 
U(1) R-symmetry, as the world-volume field strengths for the $(p,q)$ 
five-branes illustrate~\cite{Cederwall:1998a,Westerberg:1999b}:
\bea
\cF_2 &=& \cU^r \D A_{1;r} - \cC_2\,, \qquad\qquad
F_4 = \D A_3 - C_4 + \half\,\im(\cC_2\, \bcF_2)\,, \non \\
\cF_6 &=& \cU^r \D A_{5;r} - \cC_6 + x\,\cC_2\, F_4 -
(1{-}x)\,C_4\,\cF_2 +
\half(\mbox{$\frac{2}{3}$}{-}x)\,\im(\cC_2\,\bcF_2)\,\cF_2 
+\,\, \half(\third{-}x)\,\im(\cC_2\,\bcF_2)\,\cC_2\,.
\eea
These fields are all both gauge and SL(2) invariant. Whereas the four-form
field strength $F_4$ is also U(1) neutral, the complex two- and six-form field
strengths have U(1) charge $+1$, a fact which we indicate by the use of 
calligraphic letters (complex conjugation, indicated with a bar,
reverses the sign of the U(1) charge). 
Analogously, $\cC_2$, $C_4$ and $\cC_6$ are SL(2)-invariant versions of the 
supergravity gauge potentials, and $\cH_3$, $H_5$ and $\cH_7$ in
eq.~(\ref{wvbis}) below the corresponding field strengths (of which 
$\cH_7$ depends on the interpolation parameter $x$). 
The precise expressions for the world-volume field strengths were determined 
in the usual fashion by gauge invariance. They satisfy the
Bianchi identities ($\DD$ is the U(1)-covariant derivative 
and $P$ is a covariantly constant one-form of U(1) 
charge $+2$~\cite{Howe:1984})
\bea
\label{wvbis}
&&\DD \cF_2 + i\,\bcF_2\,P + \cH_3 = 0\,, \qquad\qquad
\D F_4 + H_5 + \half\,\im(\bcF_2\,\cH_3) =0\,,\non \\
&&\DD \cF_6 + i\,\bcF_6 \,P + \cH_7 - x\,\cH_3\,F_4 +
(1{-}x)\,H_5\,\cF_2+\half(\mbox{$\frac{2}{3}$}{-}x)\,\cF_2\,\im(\cF_2\,\bcH_3)
= 0\,.
\eea
The Ansatz for the action reads
\be
\label{Ansatz}
S  = \int \la \left[ {*}1 + 
{*}\Phi(\cF_2,\bcF_2,F_4) + \cF_6\,{*}\bcF_6\right]\,\,,
\ee
where $\la$ is a Lagrange multiplier for the constraint 
$\Ups = 1 + \Phi(\cF_2,\bcF_2,F_4) - {*}\cF_6\,{*}\bcF_6 \approx 0$, and 
$\Phi$ is required to have U(1) charge zero but is otherwise unconstrained.
Compatibility between the equations of motion encoded in~(\ref{Ansatz}) and 
the Bianchi identities~(\ref{wvbis}) require the duality relations to take 
the form~\cite{Cederwall:1998a} 
\be
\label{dualrel}
 -2x\,\re({*}\cF_{6}\,{*}\bcF_{2}) = K_{4} := \frac{\de\Phi}{\de
 F_4}\,,\qquad
(1{-}x)\,{*}\cF_{6}\,{*}F_{4} +
 \mbox{$\frac{i}{6}$}\,{*}[\re({*}\cF_6\,\bcF_2)\we \cF_2] =
 \cK_{2} := \frac{\de\Phi}{\de \bcF_2}\,.
\ee
The complicated structure of these relations, which can be traced back to
the fact that the tension field strength is complex, makes the $\ka$-symmetry
analysis difficult and necessitates a perturbative approach.
For an outline of this analysis and a more extensive discussion, we refer
the reader to ref.~\cite{Westerberg:1999b}, quoting here only the final
results for the projection operator and the action valid up to fourth order 
in the world-volume field strengths ($\bet$ is a free parameter in 
calculations up to this order):
\be
\label{IIB5projop}
2\,{*}\ga_6\,P_\pm\,\ze = {*}\ga_6\,\ze \mp
\big[\mbox{$\frac{2i}{3}$}\,{*}F_4\scal \ga_2\,\ze + \mbox{$\frac{i}{3}$}
\,{*}\cF_2\scal \ga_4\,\bar{\ze} + {*}\cF_6\,\bar{\ze} \big] + \cO(F^5)
\,,
\ee
\bea
\label{IIB5action}
S &=& \int\D^6\xi\,\sqrt{-g}\,\la\,\big[ 1 + \third \,\cF_2\scal \bcF_2 +
\mbox{$\frac{2}{3}$}\,F_4\scal F_4 +
\sixth(\beta{-}2)\,{*}(\cF_2\we F_4)\,{*}(\bcF_2\we F_4)
+  \sixth(\beta{-}\mbox{$\frac{2}{3}$})\,\cF_2\scal
\bcF_2 \, F_4\scal F_4 \non\\
&&\hspace{6mm}+\,\,
\sixth(1{-}\beta)\,(\cF_2\we\bcF_2)\scal({*}F_4\we{*}F_4) +
\sixth\,\beta\,(\cF_2\we{*}F_4)\scal (\bcF_2\wedge 
{*}F_4) + \cO(F^6) - {*}\cF_6\,{*}\bcF_6 \big]\,.
\eea
A somewhat peculiar feature worth noting, is that the interpolation
parameter $x$ gets fixed (to the value~$\mbox{$\frac{2}{3}$}$) in the
process.

\subsection*{Discussion}

To summarise, we have found that using a formulation which involves
auxiliary duality relations has some advantages when constructing 
world-volume actions for super-$p$-branes. One point worth mentioning
in this context, is that this formulation allows for symmetries to
be handled in a natural way, an important example being 
actions with manifest SL(2) covariance in the type-{\II}B theory. 
Another important aspect is the constructive nature of the approach:
even in the cases where the form of the bosonic action is not known,
one can determine the full action in an essentially algorithmic fashion 
using gauge invariance and $\ka$-symmetry.
The essential steps of the method are:
\begin{itemize}
\item Construct gauge-invariant world-volume field strengths and
derive their Bianchi identities.
\item Using the general Ansatz for the action, derive the associated duality 
relations compatible with the world-volume Bianchi identities.
\item Derive the $\ka$-variation of the action (or equivalently of the 
constraint $\Ups$), eliminating all direct dependence of the unknown
function $\Phi$ in the action Ansatz by means of the duality relations.
\item Make an Ansatz for the $\ka$-symmetry projection operator. 
(This is the only step which involves some guesswork; we would like to stress 
though that in all known cases the projection operator has a very natural 
form.)
\item Insert the background supergravity constraints and expand the
variation in its irreducible components. This leads to a set of constraints
which can be used to deduce the duality relations and hence also the action. 
(One can also, at least in principle, use the requirement of $\ka$-invariance 
to learn about the background supergravity constraints.)
\end{itemize}
Moreover, we would like to stress that in our approach all gauge symmetries 
are made manifest; other approaches lead to actions which do not have this 
property. Such examples are the dual D3- and D4-brane actions in 
ref.~\cite{Aganagic:1997} as well as the type-{\II}B NS5-brane action in 
ref.~\cite{Eyras:1998} (whereas for the former two cases one can show that 
there is agreement with our approach at the level of the equations of motion, 
this has not been shown for the case of the NS5-brane).

For higher-dimensional cases the actions in their most general form tend to 
become complicated. For the D-branes it is always possible to choose the
parameters so that all world-volume fields except $F_2$ decouple. In
this limit the auxiliary duality relations become redundant and can
be dropped, and the action becomes the standard one; here our construction
is thus a natural generalisation of earlier work. For other cases 
(such as the NS5-brane) there is no choice of the parameters for which
the auxiliary duality relations become superfluous. This fact seems to reflect 
an inherent obstruction to the construction of conventional actions with all
gauge symmetries manifested (cf.\ the M5-brane, where no action with a chiral
three-form exists~\cite{Witten:1996a}, but where, when supplemented with its 
anti-chiral part (`doubled') and an auxiliary duality relation, a generalised 
action of the form discussed 
above can be constructed~\cite{Cederwall:1998b}). Thus, the price one has to 
pay for having manifest symmetries is the introduction of auxiliary duality
relations.

As far as further applications of the method are concerned, one could 
proceed in various directions. For instance, 
it has been argued that there should exist certain exotic extended 1/2-BPS
objects which do not show up in Hull's brane scan~\cite{Hull:1997} 
(for some recent results and discussion, see ref.~\cite{Eyras:1999}). 
Such objects---whose masses scale with the string coupling in a non-standard
fashion as $g_{\rm s}^{-k}$ with $k>2$---are required to fill up U-duality 
multiplets for torus compactifications of M~theory in the Matrix-theory
formulation~\cite{Elitzur:1997,Hull:1998,Obers:1998}. 
In particular, U-duality considerations seem to indicate the existence of an 
M8-brane in $D=11$, which would give rise to exotic seven- and eight-branes 
in the $D=10$ type-{\II}A theory~\cite{Elitzur:1997}.
In the type-{\II}B theory, the existence of an NS7-brane whose mass goes
as $g_{\rm s}^{-3}$ is generally accepted, and one may ask whether a similar 
object might exist on the type-{\II}A side. However, it is readily seen 
that a $\ka$-symmetry projection operator (and hence a 1/2-BPS brane) can not 
be constructed in the standard manner for this case. Assuming, nevertheless, 
that this problem could somehow be circumvented, a possible candidate for a 
type-{\II}A seven-brane is suggested by the existence in the doubled 
formulation of type-{\II}A supergravity of a field strength 
$H_9 = \D B_8 + \cdots\,$, where the potential $B_8$ is dual to the 
dilaton. Indeed, one can construct a gauge-invariant world-volume field 
strength that couples to $B_8$. Although it appears difficult to 
$\ka$-symmetrise this construction (if at all possible), let us end by 
noting that for every other background supergravity potential it has been 
possible to construct an associated gauge-invariant WZ term and subsequently 
a $\ka$-symmetric world-volume action describing a brane which couples to the 
potential.

\subsection*{Acknowledgements}

The work of A.W. and N.W. was supported by the European Commission under
contracts FMBICT972021 and FMBICT983302, respectively.
A.W. would like to thank the organisers of QFTHEP'99 for the invitation to 
give a talk, and the organisers of the TMR meeting in Paris for the 
opportunity to present this work.


\providecommand{\href}[2]{#2}
\begingroup\raggedright\endgroup


\begin{thebibliography}{99}

\bibitem{Cederwall:1996b}
M.~Cederwall, A.~von Gussich, B.~E.~W. Nilsson, and A.~Westerberg, ``The
  Dirichlet super-three-brane in ten-dimensional type {\II}B supergravity.'' Nucl.
  Phys. {\bf B490} (1997) 163--178,
  \href{http://xxx.lanl.gov/abs/hep-th/9610148}{{\tt hep-th/9610148}}.

\bibitem{Cederwall:1996c}
M.~Cederwall, A.~von Gussich, B.~E.~W. Nilsson, P.~Sundell, and A.~Westerberg,
  ``The Dirichlet super-$p$-branes in ten-dimensional type {\II}A and {\II}B
  supergravity.'' Nucl. Phys. {\bf B490} (1997) 179--201,
  \href{http://xxx.lanl.gov/abs/hep-th/9611159}{{\tt hep-th/9611159}}.

\bibitem{Aganagic:1996a}
M.~Aganagic, C.~Popescu, and J.~H. Schwarz, ``D-brane actions with local kappa
  symmetry.'' Phys. Lett. {\bf B393} (1997) 311--315,
  \href{http://xxx.lanl.gov/abs/hep-th/9610249}{{\tt hep-th/9610249}}.

\bibitem{Aganagic:1996b}
M.~Aganagic, C.~Popescu, and J.~H. Schwarz, ``Gauge invariant and gauge fixed
  D-brane actions.'' Nucl. Phys. {\bf B495} (1997) 99,
  \href{http://xxx.lanl.gov/abs/hep-th/9612080}{{\tt hep-th/9612080}}.

\bibitem{Bergshoeff:1996c}
E.~Bergshoeff and P.~K. Townsend, ``Super D-branes.'' Nucl. Phys. {\bf B490}
  (1997) 145--162, \href{http://xxx.lanl.gov/abs/hep-th/9611173}{{\tt
  hep-th/9611173}}.

\bibitem{Bandos:1997}
I.~Bandos {\em et~al.}, ``Covariant action for the superfive-brane of M
  theory.'' Phys. Rev. Lett. {\bf 78} (1997) 4332--4334,
  \href{http://xxx.lanl.gov/abs/hep-th/9701149}{{\tt hep-th/9701149}}.

\bibitem{Aganagic:1997b}
M.~Aganagic, J.~Park, C.~Popescu, and J.~H. Schwarz, ``World volume action of
  the M theory five-brane.'' Nucl. Phys. {\bf B496} (1997) 191,
  \href{http://xxx.lanl.gov/abs/hep-th/9701166}{{\tt hep-th/9701166}}.

\bibitem{Duff:1993}
M.~J. Duff and J.~X. Lu, ``Type {\II} $p$-branes: the brane scan revisited.''
  Nucl. Phys. {\bf B390} (1993) 276--290,
  \href{http://xxx.lanl.gov/abs/hep-th/9207060}{{\tt hep-th/9207060}}.

\bibitem{Witten:1996a}
E.~Witten, ``Five-brane effective action in M theory.'' J. Geom. Phys. {\bf 22}
  (1997) 103, \href{http://xxx.lanl.gov/abs/hep-th/9610234}{{\tt
  hep-th/9610234}}.

\bibitem{Hull:1997}
C.~M. Hull, ``Gravitational duality, branes and charges.'' Nucl. Phys. {\bf
  B509} (1998) 216, \href{http://xxx.lanl.gov/abs/hep-th/9705162}{{\tt
  hep-th/9705162}}.

\bibitem{Obers:1998}
N.~A. Obers and B.~Pioline, ``U duality and M theory.'' Phys. Rep. {\bf 318}
  (1999) 113, \href{http://xxx.lanl.gov/abs/hep-th/9809039}{{\tt
  hep-th/9809039}}. 

\bibitem{Cederwall:1997}
M.~Cederwall and P.~K. Townsend, ``The manifestly Sl(2,$\Z$) covariant
  superstring.'' JHEP {\bf 9709} (1997) 003,
  \href{http://xxx.lanl.gov/abs/hep-th/9709002}{{\tt hep-th/9709002}}.

\bibitem{Cederwall:1998a}
M.~Cederwall and A.~Westerberg, ``World-volume fields, SL(2,$\Z$) and duality:
  the type {\II}B three-brane.'' JHEP {\bf 9802} (1998) 004,
  \href{http://xxx.lanl.gov/abs/hep-th/9710007}{{\tt hep-th/9710007}}.

\bibitem{Westerberg:1999b}
A.~Westerberg and N.~Wyllard, ``Towards a manifestly SL(2,$\Z$)-covariant
  action for the type {\II}B $(p,q)$ super-five-branes.'' JHEP {\bf 06} (1999)
  006, \href{http://xxx.lanl.gov/abs/hep-th/9905019}{{\tt hep-th/9905019}}.

\bibitem{Bergshoeff:1998c}
E.~Bergshoeff and P.~K. Townsend, ``Super D-branes revisited.'' Nucl. Phys.
  {\bf B531} (1998) 226, \href{http://xxx.lanl.gov/abs/hep-th/9804011}{{\tt
  hep-th/9804011}}.

\bibitem{Townsend:1996a}
P.~K. Townsend, ``D-branes from M-branes.'' Phys. Lett. {\bf B373} (1996)
  68--75, \href{http://xxx.lanl.gov/abs/hep-th/9512062}{{\tt hep-th/9512062}}.

\bibitem{Eyras:1998}
E.~Eyras, B.~Janssen, and Y.~Lozano, ``Five-branes, KK monopoles and T
  duality.'' Nucl. Phys. {\bf B531} (1998) 275,
  \href{http://xxx.lanl.gov/abs/hep-th/9806169}{{\tt hep-th/9806169}}.

\bibitem{Cederwall:1998b}
M.~Cederwall, B.~E.~W. Nilsson, and P.~Sundell, ``An action for the
  superfive-brane in $D{=}11$ supergravity.'' JHEP {\bf 9804} (1998) 007,
  \href{http://xxx.lanl.gov/abs/hep-th/9712059}{{\tt hep-th/9712059}}.

\bibitem{Westerberg:1999a}
A.~Westerberg and N.~Wyllard, ``Supersymmetric brane actions from interpolating
  dualisations.'' Nucl. Phys. {\bf B560} (1999) 683--715,
  \href{http://xxx.lanl.gov/abs/hep-th/9904117}{{\tt hep-th/9904117}}.

\bibitem{Callan:1991}
C.~G. {Callan Jr}, J.~A. Harvey, and A.~Strominger, ``Worldbrane actions for
  string solitons.'' Nucl. Phys. {\bf B367} (1991) 60--82.

\bibitem{Argurio:1997}
R.~Argurio, F.~Englert, L.~Houart, and P.~Windey, ``On the opening of branes.''
  Phys. Lett. {\bf B408} (1997) 151--156,
  \href{http://xxx.lanl.gov/abs/hep-th/9704190}{{\tt hep-th/9704190}}.

\bibitem{Howe:1984}
P.~S. Howe and P.~C. West, ``The complete $N=2$, $d=10$ supergravity.'' Nucl.
  Phys. {\bf B238} (1984) 181.

\bibitem{Aganagic:1997}
M.~Aganagic, J.~Park, C.~Popescu, and J.~H. Schwarz, ``Dual D-brane actions.''
  Nucl. Phys. {\bf B496} (1997) 215,
  \href{http://xxx.lanl.gov/abs/hep-th/9702133}{{\tt hep-th/9702133}}.

\bibitem{Eyras:1999}
E.~Eyras and Y.~Lozano, ``Exotic branes and nonperturbative seven-branes.''
  \href{http://xxx.lanl.gov/abs/hep-th/9908094}{{\tt hep-th/9908094}}.

\bibitem{Elitzur:1997}
S.~Elitzur, A.~Giveon, D.~Kutasov, and E.~Rabinovici, ``Algebraic aspects of
  matrix theory on $T^d$.'' Nucl. Phys. {\bf B509} (1998) 122,
  \href{http://xxx.lanl.gov/abs/hep-th/9707217}{{\tt hep-th/9707217}}.

\bibitem{Hull:1998}
C.~M. Hull, ``Matrix theory, U duality and toroidal compactifications of M
  theory.'' JHEP {\bf 10} (1998) 011,
  \href{http://xxx.lanl.gov/abs/hep-th/9711179}{{\tt hep-th/9711179}}.

\end{thebibliography}
\end{document}